\documentstyle[aps,preprint]{revtex}
\begin{document} 
\draft 
\preprint{USC(NT)-95-6, nucl-th/9512023}
\title{Chiral Perturbation Approach 
to the $pp \rightarrow pp\pi^0$
Reaction \\
Near Threshold} 

\author{B.-Y. Park,\thanks{On leave 
of absence from Department of Physics, 
Chungnam National University, 
Daejeon 305-764, Korea}  
F.Myhrer, J.R.Morones, 
T.Meissner and K.Kubodera} 

\address{Department of
Physics and Astronomy, 
University of South Carolina \\ 
Columbia, SC 29208, U. S. A.}

\maketitle

\begin{abstract}
The usual theoretical treatments of the near-threshold 
$pp \rightarrow pp\pi^0$ reaction are based on various phenomenological Lagrangians.
In this work we examine the relationship between these approaches 
and a systematic chiral perturbation method.
Our chiral perturbation calculation indicates 
that the pion rescattering term should be significantly enhanced as
compared with the traditional phenomenological treatment, 
and that this term should have substantial energy and 
momentum dependence.
An important consequence of 
this energy-momentum dependence is that, 
for a representative threshold kinematics
and within the framework of our semiquantitative calculation, 
the rescattering term interferes destructively with the Born-term
in sharp contrast to the constructive interference 
obtained in the conventional treatment. 
This destructive interference makes 
theoretical cross sections for $pp \rightarrow pp\pi^0$
much smaller than the experimental values,
a feature that suggests the importance 
of the heavy-meson exchange contributions to 
explain the experimental data.
\end{abstract}


\narrowtext
\section{Introduction}
Recently Meyer et al. \cite{meyetal90}
carried out  high-precision measurements 
of the total cross sections near threshold 
for the reaction
\begin{equation}
p+p \rightarrow p+p+\pi^0\,.
\label{eq:pppi}
\end{equation}
These measurements were confirmed 
by Bondar et al. \cite{Uppsala}.
The early theoretical calculations
\cite{kr66,ms91,nis92} underestimate 
these s-wave $\pi^0$ production 
cross sections by a factor of $\sim$5.
The basic features of these early calculations
may be summarized as follows.
The pion production reactions are 
assumed to be described by 
the single nucleon process (the Born term), 
Fig.1(a),
and the $s$-wave pion rescattering process,
Fig.1(b). 
The $\pi$-$N$ vertex for the Born term is 
assumed to be given by the 
pseudovector interaction Hamiltonian  
\begin{equation}
{\cal H}_0 = \frac{g_A}{2 f_\pi} \bar{\psi} 
\left( \bbox{\sigma} \!\cdot\! \bbox{\nabla} 
( \bbox{\tau} \!\cdot\! \bbox{\pi} )
- \frac{i}{2m_N} 
\{ \bbox{\sigma} \!\cdot\! \bbox{\nabla}, 
\bbox{\tau} \!\cdot\! \dot{\bbox{\pi}} \} 
\right) \psi, 
\label{H0}
\end{equation}
where $g_A$ is the axial coupling constant,
and $f_\pi$ = 93 MeV is 
the pion decay constant.
The first term represents 
$p$-wave pion-nucleon coupling,
while the second term accounts 
for the nucleon recoil effect and makes ${\cal H}_0$ 
``Galilean-invariant".
For $s$-wave pion production
only the second term contributes.
Since this second term is smaller 
than the first term 
by a factor of $\sim m_\pi /m_N$, 
the contribution of the Born term 
to $s$-wave pion production
is intrinsically suppressed,
and as a consequence
the process becomes sensitive 
to two-body contributions, Fig.1(b).
The s-wave rescattering vertex in Fig.1(b)
is commonly calculated 
using the phenomenological Hamiltonian \cite{kr66} 
\begin{equation}
{\cal H}_{1} = 
4\pi \frac{\lambda_1}{m_\pi} \bar{\psi} 
\bbox{\pi}\!\cdot\!\bbox{\pi} \psi
+ 4\pi \frac{\lambda_2}{m^2_\pi} 
\bar{\psi} \bbox{\tau}\!\cdot\!
\bbox{\pi} \!\times\! \dot{\bbox{\pi}} \psi
\label{H1}
\end{equation}
The two coupling constants 
$\lambda_1$ and $\lambda_2$ in Eq.(\ref{H1}) were determined 
from the $S_{11}$ and $S_{31}$ 
pion nucleon scattering lengths 
$a_{1/2}$ and $a_{3/2}$ as
\begin{mathletters}
\label{eq:4}
\begin{equation}
\lambda_1 = 
\frac{m_\pi}{6} 
\left( 1 + \frac{m_\pi}{m_N} 
\right) \,(a_{1/2} + 2 a_{3/2}), 
\label{eq4a} 
\end{equation}
\begin{equation}
\lambda_2 =  
\frac{m_\pi}{6} 
\left( 1 + \frac{m_\pi}{m_N} \right)\, 
(a_{1/2} -  a_{3/2}). 
\label{eq4b}
\end{equation}
\end{mathletters}
The current algebra prediction \cite{wei66} 
for the scattering lengths, 
$a_{1/2} = - 2a_{3/2} = 
m_\pi /4\pi f_\pi^2 = 0.175 m_\pi^{-1}$,
implies that only chiral symmetry breaking terms 
will give a non-vanishing value of 
the coupling constant $\lambda_1$ 
in Eq.(\ref{H1}).
Therefore $\lambda_1$ is expected
to be very small. 
Indeed, the empirical values 
$a_{1/2} \simeq 0.175 m^{-1}_\pi$ and 
$a_{3/2}\simeq -0.100 m^{-1}_\pi$ obtained by 
H\"{o}hler {\it et al.\/}\cite{hoe83}
lead to 
$\lambda_1 \sim 0.005$ and
$\lambda_2 \sim 0.05$. 
So the contribution of the $\lambda_1$ term 
in Eq.(\ref{H1}) is significantly suppressed.
Meanwhile, although
$\lambda_2$ is much larger than $\lambda_1$,
the isospin structure of the $\lambda_2$ term 
is such that it cannot contribute 
to the $\pi^0$ production 
from two protons at the rescattering vertex in Fig.1(b). 
Thus, the use of the phenomenological 
Hamiltonians, Eqs.(\ref{H0}) and (\ref{H1}), to calculate
the Born term and the rescattering terms illustrated in Figs.1(a)
and (b), gives significantly suppressed cross sections for 
the $pp \rightarrow pp\pi^0$ reaction
near threshold. 
Therefore, theoretically calculated  cross sections 
can be highly sensitive to 
any deviations from this conventional treatment.
These delicate features should be kept in mind 
in discussing the large discrepancy 
(a factor of $\sim$5)
between the observed cross sections 
and the predictions of the earlier calculations.

A plausible mechanism 
to increase the theoretical cross section 
was suggested by Lee and Riska\cite{lr93}. 
They proposed to supplement 
the contribution of the pion-exchange diagram,
Fig.1(b),
with the contributions 
of the short-range axial-charge 
exchange operators which were directly 
related to heavy-meson exchanges in the
nucleon-nucleon interactions \cite{br92}.
According to Lee and Riska,
the shorter-range meson exchanges 
(scalar and vector exchange contributions)
can enhance the cross section by a factor  3--5. 
Subsequently, Horowitz {\it et al.} \cite{hmg94}
demonstrated, for the Bonn meson exchange potential, 
a prominent role of the $\sigma$ meson
in enhancing the cross section, 
thereby basically confirming 
the conclusions of Lee and Riska. 
The possible importance of heavy-meson
exchanges may be inferred 
from the following simple argument.
Consider Fig.1(b) in the center-of-mass (CM) system
with the initial and final interactions turned off 
and with the exchanged particle allowed to be 
any particle (not necessarily a pion). 
At threshold,
$q_0 = m_\pi$, $\bbox{q} = 0$, 
$\bbox{p}_1'=\bbox{p}_2'=0$,
so that any exchanged particle must have
$k_0 = m_\pi/2= 70$ MeV and 
$|\bbox{k}| = \sqrt{m_\pi m_N + (m_\pi /2)^2} \sim 370$ MeV/c,
which implies
$k^2 = -m_\pi m_N$.
Thus the rescattering process 
probes two-nucleon forces at distances $\sim$ 0.5 fm  
corresponding to a typical effective exchanged mass 
$\sqrt{m_\pi m_N}\,=\,370$ MeV.
Its sensitivity to the intermediate-range  $N$-$N$ forces 
indicates the possible importance of the
two-body heavy meson axial exchange currents
considered by Lee and Riska.
The particular kinematical situation we considered here
shall be referred to as the {\it typical threshold kinematics}.

Meanwhile, Hern\'{a}ndez and Oset \cite{ho95} 
considered the {\em off-shell} 
dependence of the $\pi N$ $s$-wave isoscalar 
amplitude featuring 
in the rescattering process, Fig.1(b). 
They pointed out that 
the $s$-wave amplitude could be appreciably enhanced 
for off-shell kinematics
pertinent to the rescattering process. 
We have seen above that, for the {\it typical threshold kinematics},
the exchanged pion can indeed be far off-shell.
The actual kinematics of course 
may deviate from the {\it typical threshold kinematics}
rather significantly due to energy-momentum exchanges 
between the two nucleons in the initial and final states,
but the importance of the off-shell kinematics
for the exchanged pion is likely to persist.
Hern\'{a}ndez and Oset
examined two types of off-shell extrapolation: 
(i) the Hamilton model 
for $\pi N$ isoscalar amplitude 
based on $\sigma$-exchange plus 
a short range piece\cite{ham67}, and 
(ii) an extrapolation based 
on the current algebra constraints.
In either case the enhancement 
of the total cross section due to 
the rescattering process 
was estimated to be strong enough 
to reproduce the experimental data. 
A more detailed momentum-space calculation 
carried out by Hanhart {\it et al.}\cite{hanetal95} 
supports the significant enhancement 
due to an off-shell effect 
in the rescattering process,
although the enhancement is not large enough 
to explain the experimental data.
It should be emphasized 
that Hanhart {\it et al.}'s calculation 
eliminates many of 
the kinematical approximations 
employed in the previous calculations.

Given these developments based on
the phenomenological Lagrangians,
we consider it important to examine 
the significance of 
these phenomenological Lagrangians
in chiral perturbation theory
($\chi$PT) \cite{gl84,bkm95}
which in general serves as
a guiding principle
for low-energy hadron dynamics.
In the present work 
we shall describe an attempt at
relating the traditional phenomenological approaches
to $\chi$PT.
The fact that $\chi$PT accounts for and improves 
the results of the current algebra
also makes it a natural framework 
for studying threshold pion production. 
Furthermore, in this low-energy regime,
it is natural to employ 
the heavy-fermion formalism (HFF)\cite{jm91}.
The HFF has an additional advantage 
of allowing easy comparison with
Eqs.(\ref{H0}) and (\ref{H1}).

It should be mentioned, however, 
that the application of $\chi$PT to nuclei
involves some subtlety. 
As emphasized by Weinberg \cite{wei90},
naive chiral counting
fails for a nucleus, 
which is a loosely bound many-body system.
This is because purely nucleonic intermediate states
occurring in a nucleus
can have very low excitation energies,
which spoils the ordinary chiral counting.
To avoid this difficulty,
one must first classify diagrams
appearing in perturbation series
into irreducible and reducible diagrams, 
according to whether or not a diagram is 
free from purely nucleonic intermediate states.
Thus, in an irreducible diagram,
every intermediate state 
contains at least one meson.
The $\chi$PT can be safely applied 
to the irreducible diagrams.
The contribution of 
all the irreducible diagrams 
(up to a specified chiral order)
is then to be used as an effective operator
acting on the nucleonic Hilbert space.
This second step allows us 
to incorporate the contributions of 
the reducible diagrams.
We may refer to this two-step procedure 
as the {\it nuclear chiral perturbation theory} 
(nuclear $\chi$PT).
This method was first applied by Weinberg \cite{wei90} to
chiral-perturbation-theoretical derivation 
of the nucleon-nucleon interactions
and subsequently used by van Kolck {\it et al.} \cite{kol92}.
Park, Min and Rho  (PMR) \cite{pmr93} 
applied the nuclear $\chi$PT
to meson exchange currents in nuclei.
The success of the nuclear $\chi$PT
in describing the exchange currents
for the electromagnetic and weak interactions
is well known \cite{pmr93,ptk94,pmr95}.
The present paper
is in the spirit of the work of PMR.

This article is organized as follows: 
In the next section we define our pion field and the chiral 
counting procedure.
Then in section III we present the two lowest order Lagrangians,
discuss their connection to the early works on this reaction
and determine within certain approximations the numerical values of 
the effective pion rescattering vertex strength, $\kappa_{th}$.
In section IV we briefly discuss the connection between the transition 
matrix for this reaction and the $\chi$PT calculated amplitude.
In section V we present necessary loop corrections to the Born term,
and in section VI we calculate the cross section and discuss the various
approximations and the uncertainties of the low 
energy constants in $\chi$PT.
Finally in section VII, after discussing
some higher chiral order diagrams, we present our main
conclusions.

A work very similar in spirit to ours
has recently been completed by Cohen 
{\it et al.}\cite{cfmv95}.

\section{Chiral Perturbation Theory}
The effective chiral Lagrangian
${\cal L}_{\rm{ch}}$ involves an SU(2) matrix $U(x)$
that is non-linearly related to the pion field
and that has standard chiral 
transformation properties \cite{geo84}. 
An example is \cite{com1}
\begin{equation}
U(x) = \sqrt{1-[\bbox{\pi}(x)/f_\pi]^2}
+i\bbox{\tau} \!\cdot\! \bbox{\pi}(x)/f_\pi.
\label{eq:UUU}
\end{equation}
In the meson sector,
the sum of chiral-invariant monomials 
constructed from $U(x)$ and its derivatives 
constitutes the chiral-symmetric part of
${\cal L}_{\rm{ch}}$.
Furthermore, 
one can construct systematically 
the symmetry-breaking part of ${\cal L}_{\rm{ch}}$
with the use of a mass matrix ${\cal M}$ 
the chiral transformation of which is
dictated by that of the quark mass term 
in the QCD Lagrangian. 
To each term appearing in ${\cal L}_{\rm{ch}}$
one can assign a chiral order index $\bar{\nu}$ 
defined by 
\begin{equation}
\bar{\nu} \equiv d -2,
\label{eq:nubar1}
\end{equation}
where $d$ is the summed power of
the derivative and the pion mass involved in this term.
A low energy phenomenon is characterized 
by a generic pion momentum $Q$, 
which is small compared to the chiral scale 
$\Lambda \sim $  1 GeV.
It can be shown that 
the contribution of a term
of  chiral order $\bar{\nu}$ 
carry a factor 
$(\tilde{Q}/\Lambda)^{\bar{\nu}}$,
where $\tilde{Q}$
represents either $Q$ or 
the pion mass $m_\pi$.
This suggests the possibility
of describing low-energy phenomena
in terms of ${\cal L}_{\rm{ch}}$
that contains only 
a manageably limited number 
of terms of  low chiral order. 
This is the basic idea of $\chi$PT.

The heavy fermion formalism
(HFF)\cite{jm91} allows us 
to easily extend $\chi$PT to the 
meson-nucleon system.
In HFF,
the ordinary Dirac field $\psi$
describing the nucleon, 
is replaced by the heavy nucleon field $N(x)$ 
and the accompanying ``small component field"
$n(x)$ through the transformation

\begin{equation}
\psi(x)
 = \exp(-i m_N v\cdot x)\,
[N(x) + n(x)]
\label{N} 
\end{equation}
with
\begin{equation}
/\hskip -0.52em{v} N = N, \hskip 1em 
/\hskip -0.52em{v} n = - n,  
\end{equation}
where the four-velocity $v_\mu$ is
assumed to be almost static, i.e.,
$v_\mu\approx(1,0,0,0)$ \cite{com2}.
Elimination of $n(x)$ in favor of $N(x)$
leads to expansion in $\partial_\mu/m_N$.
Since $m_N\approx$ 1 GeV $\approx\Lambda$,
an expansion in $\partial_\mu/m_N$ 
may be treated like 
an expansion in $\partial_\mu/\Lambda$.
${\cal L}_{\rm{ch}}$ in HFF
consists of chiral symmetric monomials 
constructed from $U(x)$, $N(x)$ and
their derivatives
and of symmetry-breaking terms 
involving ${\cal M}$.
The chiral order $\bar{\nu}$ in HFF 
is defined by 
\begin{equation}
\bar{\nu} \equiv d + n/2 -2,
\label{eq:nubar2}
\end{equation} 
where $d$ is, as before,
the summed power of
the derivative and the pion mass, 
while $n$ is the number of nucleon fields 
involved in a given term.
As before, a term in ${\cal L}_{\rm{ch}}$ 
with chiral order $\bar{\nu}$
can be shown to carry a factor 
$(\tilde{Q}/\Lambda)^{\bar{\nu}} \ll 1$.
In what follows, $\bar{\nu}$ stands for
the chiral order defined in Eq.(\ref{eq:nubar2}).

In addition to the chiral order index $\bar{\nu}$
defined for each term in ${\cal L}_{\rm{ch}}$,
we assign a chiral order index $\nu$ 
for each irreducible Feynman diagram
appearing in the chiral perturbation series
for a multifermion system \cite{wei90}.
Its definition is 
\begin{equation}
\nu = 4 - E_N - 2C + 2L + \sum_i \bar{\nu}_i,
\label{eq:nununu}
\end{equation}
where $E_N$ is the number of 
nucleons in the Feynman diagram, 
$L$ the number of loops, 
and $C$ the number of disconnected parts
of the diagram.
The sum over $i$ runs over all the vertices 
in the Feynman graph, 
and $\bar{\nu}_i$ is the chiral order 
of each vertex. 
One can show \cite{wei90} that
an irreducible diagram of chiral order $\nu$
carries a factor 
$(\tilde{Q}/\Lambda)^{\nu} \ll 1$. 

In the literature 
the term ``effective Lagrangian" 
(or ``effective Hamiltonian") is
often used to imply that
that Lagrangian (or Hamiltonian)
is only meant for calculating tree diagrams.
The Hamiltonians given in 
Eqs.(\ref{H0}) and (\ref{H1})
are regarded as effective Hamiltonians
of this type.
We must note, however, that
the effective Lagrangian in $\chi$PT
has a different meaning. 
Not only can ${\cal L}_{\rm{ch}}$ be used 
beyond tree approximation
but, in fact, a consistent chiral 
counting even demands inclusion of 
every loop diagram whose chiral order $\nu$
is lower than or equal to the chiral order 
of interest. 
As will be discussed below,
for a consistent $\chi$PT treatment of 
the problem at hand,
we therefore need to consider loop corrections.
However, since the inclusion 
of the loop corrections
is rather technical, 
we find it useful to first concentrate 
on the tree-diagram contributions.
This simplification allows us to 
understand the basic aspects of the relation 
between the contributions from $\chi$PT
and the phenomenological Hamiltonians,
Eqs.(\ref{H0}) and (\ref{H1}).
Therefore, in the next two sections 
(III and IV)  we limit our discussion to
tree diagrams.
A more elaborate treatment 
including loop corrections 
will be described in section V.

\section {Tree Diagram Considerations}
In order to produce the one-body and
two-body diagrams depicted in Figs.1(a)
and 1(b),
we minimally need (see below)
terms with 
$\bar{\nu}=1$ and $2$ in ${\cal L}_{\rm{ch}}$.
We therefore work with 
\renewcommand{\theequation}{\arabic{equation}}
\begin{equation}
{\cal L}_{\rm{ch}} = {\cal L}^{(0)} + {\cal L}^{(1)},
\label{Lag} 
\end{equation}
where ${\cal L}^{(\bar{\nu})}$ represents
terms of chiral order $\bar{\nu}$.
Their explicit forms are
\cite{bkm95,com3}
\begin{mathletters}
\label{eq:11}
\begin{eqnarray}
{\cal L}^{(0)} = 
&
 \frac{f^2_\pi}{4} \mbox{Tr} 
[ \partial_\mu U^\dagger \partial^\mu U 
 + m_\pi^2 (U^\dagger +  U - 2) ]  
\\ &
 + \bar{N} ( i v \cdot D + g_A^{} S \cdot u ) N 
\\ &
 - \frac12 \displaystyle \sum_A 
      C_A (\bar{N} \Gamma_A N)^2 
\label{eq17a}
\end{eqnarray}
\begin{eqnarray}
{\cal L}^{(1)} = 
&
-\frac{i g_A^{}}{2m_N} \bar{N} 
\{ S \!\cdot\! D, v \!\cdot\! u \} N 
\\ &
 + 2c_1 m_\pi^2 \bar{N} N \mbox{Tr} 
( U + U^\dagger - 2 ) 
\\ &
 + (c_2 \!-\! \frac{g_A^2}{8m_N}) \bar{N} 
(v \!\cdot\! u)^2 N 
 + c_3 \bar{N} u \!\cdot\! u  N  
\\ &
  - \frac{c_9}{2m_N} (\bar{N}N)
(\bar{N} i S \!\cdot\! u N ) 
\\ &
  - \frac{c_{10}}{2m_N} (\bar{N} S^\mu N) 
(\bar{N} i u_\mu N) 
\label{eq17b}
\end{eqnarray} 
\end{mathletters}

In the above
\begin{equation}
\xi\equiv \sqrt{U(x)},
\label{U} 
\end{equation}

\begin{equation}
 u_\mu \equiv i (\xi^\dagger \partial_\mu \xi
                - \xi \partial_\mu \xi^\dagger), 
\end{equation}

\begin{equation}
 D_\mu N \equiv (\partial_\mu
   + \textstyle\frac12 [ \xi^\dagger, 
\partial_\mu \xi] ) N,
\end{equation}
and $S_\mu$ is the covariant spin operator defined by 
\begin{equation}
S_\mu \equiv \textstyle 
\frac14 \gamma_5 
[\slash\hskip -0.52em v, \gamma_\mu].
\label{S} 
\end{equation}
In ${\cal L}^{(1)}$ above
we have retained only terms 
of direct relevance for our discussion.
The coupling constants $c_1,c_2$ and $c_3$ 
can be fixed from phenomenology \cite{bkm95}.
They are related to the pion-nucleon 
$\sigma$-term, $\sigma_{\pi N}(t) 
\sim \langle p^\prime |
\bar{m} (\bar{u}u + \bar{d}d) |p\rangle$ 
($\bar{m}=$ average mass 
of the light quarks, $t=(p^\prime - p)^2$), 
the axial polarizability $\alpha_A$ 
and the isospin-even $\pi N$ $s$-wave 
scattering length 
$a^+\equiv \frac13(a_{1/2} + 2 a_{3/2})
\approx -0.008 m_\pi^{-1}$ \cite{hoe83}. 
(The explicit expressions will be given below.) 
It should be noted that in HFF, 
a part of the term in ${\cal L}^{(1)}$ 
with the coefficient
$(c_2-g_A^2/8 m_N)$, namely      
the  $-g_A^2/8 m_N$ piece,  
represents the $s$-wave $\pi$-$N$ scattering contribution,
which in a traditional calculation is obtained from the 
crossed Born-term . 

The four-Fermi non-derivative contact terms 
in Eq.(\ref{eq:11}) were introduced 
by Weinberg \cite{wei90} and 
further investigated in two- and three-nucleon systems 
by van Kolck {\it et al}. \cite{kol92}.
Although these terms are important
in the chiral perturbative derivation of
the nucleon-nucleon interactions 
\cite{wei90,kol92},
they do not play a major role 
in the following discussion of
the threshold $pp \rightarrow pp\pi^0$ reaction.
We therefore temporarily ignore 
these four-fermion terms 
and come back to a discussion of these 
terms in the last section.

The Lagrangian (\ref{Lag}) 
leads to the pion-nucleon interaction Hamiltonian

\begin{equation}
{\cal H}_{int} = {\cal H}^{(0)}_{int} 
+ {\cal H}^{(1)}_{int}, 
\label{Hint} 
\end{equation}
where 
\begin{mathletters}
\label{eq:17}
\begin{equation}
 {\cal H}^{(0)}_{int} = 
\frac{g_A}{2f_\pi} \bar{N} 
[ \bbox{\sigma}\!\cdot\!\bbox{\nabla} 
 ( \bbox{\tau}\!\cdot\!\bbox{\pi} ) ] N 
 + \frac{1}{4f_\pi^2}
\bar{N} \bbox{\tau}\!\cdot\!\bbox{\pi}
\!\times\!\dot{\bbox{\pi}} N
\label{eq:Hint0}
\end{equation}
\begin{eqnarray}
{\cal H}^{(1)}_{int} &=&
 \frac{-i g_A}{4m_N f_\pi} \bar{N} 
\{ \bbox{\sigma}\!\cdot\!\bbox{\nabla}, 
\bbox{\tau}\!\cdot\!\dot{\bbox{\pi}} \} N  
\nonumber\\  
&& + \frac{1}{f^2_\pi} [ 2c_1 m_\pi^2 \pi^2 
 \!-\! (c_2 \!-\! \frac{g^2_A}{8m_N}) \dot{\pi}^2
 \!-\! c_3 (\partial \pi)^2 ] \bar{N} N \nonumber \\
&&
\label{eq:Hint1}
\end{eqnarray}
\end{mathletters} 
Here ${\cal H}^{(\bar{\nu})}_{int}$
represents the term of chiral order
$\bar{\nu}$.

We now compare ${\cal H}_{int}$ 
resulting from $\chi$PT, Eq.(\ref{Hint})
with the phenomenological effective Hamiltonian
${\cal H}_0+{\cal H}_1$,
Eqs.(\ref{H0}) and (\ref{H1}).
(The reader is reminded 
that the chiral index $\bar{\nu}$
should not be confused with the suffix
appearing in ${\cal H}_0$ and ${\cal H}_1$.) 
Regarding the $\pi NN$ vertices,
we note that the first term in ${\cal H}^{(0)}$
and the first term in ${\cal H}^{(1)}$
exactly correspond to the first and second terms, 
respectively, in ${\cal H}_0$.
Thus the so-called Galilean-invariance term
naturally arises as a 
$1/m_N$ correction term in HFF.
As for the $\pi\pi NN$ vertices,
we can associate the second term in $ {\cal H}^{(0)}_{int}$
to the $\lambda_2$ term in ${\cal H}_1$,
and second term in $ {\cal H}^{(1)}_{int}$
to the $\lambda_1$ term in ${\cal H}_1$.
This suggests the following identifications:
\begin{equation}
4\pi\frac{\lambda_2}{m_\pi^2}
=\frac{1}{4f_\pi^2}
\label{eq:lambda2a}
\end{equation}
and 
\begin{equation}
\begin{array}{l}
4\pi \lambda_1/m_\pi 
=
\frac{m_\pi^2}{f_\pi^2} 
[2c_1 - (c_2 - \frac{g_A^2}{8m_N})
\frac{\omega_q \omega_k}{m_\pi^2} 
 - c_3 \frac{q\cdot k}{m_\pi^2} ] 
\equiv \kappa(k,q)
\end{array}
\label{eq:kappakq}
\end{equation} 
In Eq.(\ref{eq:kappakq}),
$q=(\omega_q, {\bf q})$ and 
$k=(\omega_k, {\bf k})$
stand for the four-momenta of the
exchanged-and final pions, respectively, 
see Fig.1(b).
Since, as already discussed,
the $\lambda_2$ term is not important 
for our purposes,
we shall concentrate on the $\lambda_1$ term.
The best available estimates 
of the coefficients $c_i$ (i=1-3)
can be found in Refs.\cite{bkm95,bkm93},
which give 

\begin{mathletters}
\label{eq:20}
\begin{eqnarray}
c_1 
& 
 = - \frac{1}{4m_\pi^2} 
\left[ 
\sigma_{\pi N}(0) 
 + \frac{9 g_A^2 m_\pi^3}{64\pi f_\pi^2} 
\right] 
\\ & 
 = -0.87 \pm 0.11 \,\mbox{GeV}^{-1},
\label{C1}
\end{eqnarray}
\begin{eqnarray}
c_3 
& \displaystyle 
= - \frac{f_\pi^2}{2} \left[ \alpha_A + 
\frac{g_A^2 m_\pi}{8f_\pi^2} 
 ( \textstyle \frac{77}{48} + g_A^2 ) \right] 
\\  & 
= -5.25 \pm 0.22 \,\mbox{GeV}^{-1}, 
 \label{C2}
\end{eqnarray}
\begin{eqnarray}
c_2 
& 
= \frac{f_\pi^2}{2m_\pi^2} 
\left[ 4\pi ( 1 + \frac{m_\pi}{m_N}) a^+ 
- \frac{3g_A^2 m_\pi^3}{64\pi f_\pi^4} \right] 
\\ & \hskip 4em 
 + 2c_1 - c_3 + \frac{g_A^2}{8m_N} 
\\ & 
= 3.34 \pm 0.27 \,\mbox{GeV}^{-1}.
\label{C3}
\end{eqnarray}
\end{mathletters}
The numerical results are based on
the experimental values:
$\sigma_{\pi N}(0) = 45 \pm 8$ MeV\cite{gls91},
$\alpha_A = 2.28 
\pm 0.10 m_\pi^{-3}$\cite{hoe83},
and $a^+ =( -0.83 \pm 0.38)\cdot 10^{-2} 
m_\pi^{-1}$\cite{koc86}.
We shall show in section VI that the uncertainties
in the numerical value for $c_2$ might be
larger than quoted in Eq.(\ref{C3}).
In fact, the terms in Eqs.(\ref{C1})-(\ref{C3})
proportional to the $(g_A/f_\pi)^2$ are
${\cal O} ((m_\pi /\Lambda)^3)$ corrections 
arising from 
%{\it i.e.}, the coupling constants are not only renormalized by 
%${\cal L}^{(2)}$ counter terms   
%but also include 
finite terms of ${\cal L}^{(2)}$. 
However, since the present section is just an 
introduction to a later systematic treatment, 
this inconsistency in  ``accuracy" will be ignored
for the moment.

Now, for {\it on-shell} low energy pion-nucleon scattering,
{\it i.e.\/}, $k \!\sim\! q \!\sim\! 
(m_\pi, \bbox{0})$, we equate
\renewcommand{\theequation}
{\arabic{equation}}
\begin{equation}
4\pi\lambda_1/m_\pi
= \kappa_0\,\equiv\,
\kappa(k\!=\!(m_\pi, \bbox{0}),
q\!=\!(m_\pi, \bbox{0})),
\label{eq:onshell}
\end{equation}
where
\begin{eqnarray}
\kappa_0&=&\frac{m_\pi^2}{f_\pi^2}
\left(\tilde{c}+\frac{g_A^2}{8m_N} \right),
\label{eq:kappa}\\
\tilde{c}&\equiv& 2c_1-c_2-c_3,
\label{eq:ctilde}
\end{eqnarray}
From Eq.(\ref{eq:20}) we have
\begin{equation}
\tilde{c}\,=\,-\frac{f_\pi^2}{2m_\pi^2} 
\left[ 4\pi ( 1 + \frac{m_\pi}{m_N}) a^+ 
- \frac{3g_A^2 m_\pi^3}{64\pi f_\pi^4} \right] 
-\frac{g_A^2}{8m_N},
\label{eq:cctilde1}
\end{equation}
which results in 
\begin{equation}
\kappa_0\,=\,-2\pi
\left( 1+\frac{m_\pi}{m_N} \right)a^+ +
\frac{3g_A^2}{128\pi}
\frac{m_\pi^3}{f_\pi^4}.
\label{eq:kappa1}
\end{equation}
The above cited empirical value for $a^+$
leads to 
\begin{eqnarray}
\tilde{c} &=&(0.59\pm0.09)\,
{\rm GeV}^{-1}
\label{eq:ctilde2}\\
\kappa_0&=&(0.87\pm0.20)\,{\rm GeV}^{-1}.
\label{eq:kappa2}
\end{eqnarray}
We now interpret these results in terms of
$\lambda_1$ of Eq.(\ref{H1}).
Conventionally, $\lambda_1$ is determined
from Eq.(\ref{eq4a}) which is the first 
term in Eq.(\ref{eq:kappa1}).  Thus
\begin{equation}
\frac{4\pi\lambda_1}{m_\pi}
\,=\,-2\pi\left( 1+\frac{m_\pi}{m_N} 
\right)a^+,
\label{eqll}
\end{equation}
which gives 
\begin{equation}
\frac{4\pi\lambda_1}{m_\pi}\,=\,
(0.43\pm0.20)\,{\rm GeV}^{-1},
\label{lambdaold}
\end{equation}
or $\lambda_1\,=\,0.005\pm0.002$. 
This is the ``standard value" 
used in the literature \cite{hoe83,com4}.
On the other hand,
the r.h.s. of Eq.(\ref{eq:onshell})
based on $\chi$PT gives from Eq.(\ref{eq:kappa2})
\begin{equation}
\frac{4\pi\lambda_1}{m_\pi}\,=\,
(0.87\pm0.20)\,{\rm GeV}^{-1},
\label{eq:lambda1c}
\end{equation}
which is about twice as large as the conventional value. 
This means the second term 
in Eq.(\ref{eq:kappa1}) is almost as large
as the first term.
Thus $\chi$PT leads to a substantial
modification of the commonly used formula,
Eq.(\ref{eq4a}) or Eq.(\ref{eqll}).
This large ``higher chiral order"
corrections due to ${\cal L}^{(2)}$ 
[the term proportional to $(g_A/f_\pi)^2$
in Eq.(\ref{eq:kappa1})] 
indicates that $\chi$PT does not
converge very rapidly in this particular case. 
This apparent lack of convergence 
is probably due to the fact that
the first terms in expansion, the $\pi$-$N$
isoscalar scattering length $a^+$, 
is exceptionally small.

To develop further the connection between the traditional and the
$\chi$PT approaches, we return 
to a discussion of Eq.(\ref{eq:kappakq}).
Obviously,  the constant $\lambda_1$
cannot be fully identified with $\kappa(k,q)$ which 
depends on the momenta
$q$ and $k$.
In fact, the momentum dependence of $\kappa(k,q)$ 
should play a significant role
in describing the physical pion-nucleon elastic scattering process
where $\omega_q=\sqrt{m_\pi^2+{\bf q}^2}$,
$\omega_k=\sqrt{m_\pi^2+{\bf k}^2}$.
An additional crucial point 
in the present context is that,
in the rescattering diagram Fig.1(b),
the exchanged pion can be far off-shell, 
and therefore 
the $q$ and $k$ dependence in $\kappa(k,q)$
may play an even more pronounced role. 
As an illustration,
let us consider again the 
{\it typical threshold kinematics} 
discussed in the introduction:
$q \! \sim \! (m_\pi,\bbox{0})$ and 
$k \!\sim\! (\frac12 m_\pi, \sqrt{m_\pi m_N})$. 
If we denote by $\kappa_{th}$
the value of $\kappa(k,q)$ [Eq.(\ref{eq:kappakq})]
corresponding to the 
{\it typical threshold kinematics}, 
we have
\begin{equation}
\kappa_{th}=\frac{m_\pi^2}{f_\pi^2} 
\left[ 2c_1 - \frac{1}{2}
\left(c_2 - \frac{g^2_A}{8m_N}\right) -\frac{c_3}{2}.
\right]
\label{eq:kappath}
\end{equation}
The use of the central values 
for the coupling constants $c_1,c_2$ and $c_3$
leads to 
\begin{equation}
4 \pi \lambda_1/m_\pi = \kappa_{th}   
\sim -1.5 {\rm GeV}^{-1}.
\label{kappath1}
\end{equation}
Thus the strength of the 
$s$-wave pion-nucleon interaction here
is much stronger than the on-shell cases,
see Eqs.(\ref{lambdaold}) and (\ref{eq:lambda1c}),
and the sign of the off-shell coupling strength 
is {\it opposite} to the on-shell cases.
The first feature is qualitatively in line with 
the observation of Hern\'{a}ndez and Oset\cite{ho95}
that the rescattering term should be larger than 
previously considered.
However, the sign of the typical off-shell coupling 
in our case [Eq.(\ref{kappath1})] is
opposite to the one used in Ref.\cite{ho95}. 
As will be discussed later, this flip of the sign 
drastically changes the pattern of interplay between the Born 
and rescattering terms.
We must emphasize that the off-shell enhancement 
depends strongly on the values 
of $c_1,c_2$ and $c_3$,
which, as discussed in Ref.\cite{bkm95,bkm93}, 
are not known very accurately. 
It is therefore important to examine 
to what extent the existing large 
ambiguities in  $c_1,c_2$ and $c_3$ 
affect the off-shell enhancement of
the $pp \rightarrow pp\pi^0$ reaction.
We shall address this question in section VI.

\section{Transition Operators for 
$\lowercase{p p} \rightarrow \lowercase{p p}\pi^0$}
As explained earlier,
in the nuclear $\chi$PT
we first use $\chi$PT
to calculate the contributions
of the irreducible diagrams.
Let ${\cal T}$ represent the contributions of
all irreducible diagrams 
(up to a specified chiral order $\nu$)
for the $pp \rightarrow pp\pi^0$ process.
Then we use ${\cal T}$ 
as an effective transition operator 
in the Hilbert space of nuclear wavefunctions.
Consequently, the two-nucleon transition matrix element $T$
for the $pp \rightarrow pp\pi^0$ process
is given by
 
\begin{equation}
T\,=\,\langle \Phi_f | {\cal T} | \Phi_i \rangle,
\label{eq:Tmatrix}
\end{equation}
where $|\Phi_i\rangle$ ($|\Phi_f\rangle$)  
is the initial (final) two-nucleon state
distorted by the initial-state (final-state) interaction.
These distorted waves 
should be obtained by solving the Schr\"{o}dinger equation 
with nucleon-nucleon interactions generated
by irreducible diagrams 
pertinent to nucleon-nucleon scattering,
thereby incorporating an infinite number 
of ``reducible" ladder diagrams.
In this section we concentrate on
the derivation of the transition operator ${\cal T}$,
relegating the discussion of $T$ 
and  Eq.(\ref{eq:Tmatrix})
to section VI.

We decompose ${\cal T}$ as

\begin{equation}
{\cal T}\,=\,\sum_{\nu} {\cal T}^{(\nu)},
\label{eq:calT}
\end{equation}
where ${\cal T}^{(\nu)}$ represents
the contribution from Feynman diagrams of chiral order $\nu$,
as defined in Eq.(\ref{eq:nununu}).
The lowest value of $\nu$ 
%($\nu=-1$)
occurs for the Born term shown in Fig.2(a).
For $s$-wave $\pi$ production
at threshold the $NN\pi$ vertex 
with $\bar{\nu}=0$, the first term in Eq.(\ref{Hint}),
cannot contribute; 
hence the lowest $\bar{\nu}$
for $NN\pi$ vertex involving an external pion
must be $\bar{\nu}=1$.
The first term in Eq.(\ref{eq:Hint1})
provides this vertex.
According to Eq.(\ref{eq:nununu}),
the chiral order of Fig.2(a) is given by 
$\nu = 4-2-2\cdot 2+2\cdot 0+1\,=\,-1.$
As can be checked easily,
there are no diagrams with $\nu\,=\,0$ 
since in the rescattering diagram, Fig.2(b), 
the second term in Eq.(\ref{eq:Hint0}),
which gives the $NN\pi\pi$ vertex with $\bar{\nu}=0$,
is not operative here 
due to the isospin selection rule.
The rescattering diagram in Fig.2(b)
with the indicated value of $\bar{\nu}$
at each vertex contributes to ${\cal T}^{(\nu=1)}$.
It should be noted 
that because of the $-2C$ term 
in the chiral counting expression, Eq.(\ref{eq:nununu}),
exchange-current-type diagrams
such as Fig.2(b) 
give higher values of $\nu$.
In this work we truncate the calculation of 
the transition operator ${\cal T}$ at $\nu=1.$ 
Thus,

\begin{equation}
{\cal T}\,=\, {\cal T}^{(-1)}+{\cal T}^{(1)}
\label{eq:calTtrunc}
\end{equation}

The above enumeration is, 
as briefly discussed in section III, far from complete
because loop diagrams and counter terms and finite terms 
from ${\cal L}^{(2)}$ have been left out.
In Fig.3 we show  the loop corrections 
to the Born term [Fig.2(a)].
The diagrams in Fig.3 all have $\nu\,=\,1$
and hence are of the same chiral order 
as the leading order rescattering diagram, Fig.2(b). 
As discussed earlier, 
for the $pp \rightarrow pp\pi^0$ reaction at threshold
the contribution of the Born term
is numerically suppressed so that 
the rescattering diagram,
which is formally of higher chiral order by two units
of $\nu$, plays an essential role.
This implies that a meaningful 
and consistent $\chi$PT calculation
of this reaction must include the loop corrections 
to the leading-order Born term.
However, we continue to postpone the discussion
of loop corrections to the next section.

The tree diagrams contributing to 
Eq.(\ref{eq:calTtrunc}), Figs.2(a) and 2(b),
are as follows.
The Born term, Fig.2(a), contributes 
to ${\cal T}^{(-1)}$
and the rescattering term, Fig.2(b), contributes  
to ${\cal T}^{(1)}$. These contributions are given, respectively, by 

\begin{mathletters}
\label{eq:35}
\begin{equation}
{\cal T}^{\mbox{\scriptsize Born}}_{-1}
 = \frac{g_A}{4m_N f_{\pi}} \omega_q \sum_{i=1,2} 
\bbox{\sigma}_i \!\cdot\! 
(\bbox{p}^\prime_i + \bbox{p}_i ) {\tau}_i^0, 
\label{Tm1} 
\end{equation}
\begin{equation}
{\cal T}^{\mbox{\scriptsize Res}}_{+1}
 = -\frac{g_A}{f_\pi} \sum_{i=1,2} \kappa(k_i,q) 
\frac{\bbox{\sigma}_i \!\cdot\! \bbox{k}_i \tau_i^0 }
      { k_i^2 - m_\pi^2 + i\varepsilon}
\label{Tp1}
\end{equation}
\end{mathletters}
where $\bbox{p}_i$ and $\bbox{p^\prime}_i$ ($i=1,2$)
denote the initial and final 
momenta of the $i$-th proton,
$\bbox{k}_i \equiv \bbox{p}_i - \bbox{p^\prime}_i$;
and $\kappa(k_i,q)$ is as defined 
in Eq.(\ref{eq:kappakq}).

\section{Loop Diagrams}

We have emphasized above
that the loop corrections 
to the Born diagram, Fig.2(a),
which has chiral order $\nu=-1$, 
are of the same chiral order $\nu = 1$ as 
the two-body pion rescattering process, Fig.2(b).
These loop corrections therefore must be
included in a consistent $\nu=1$ calculation.

For our present purposes 
it is not necessary to go into a general 
discussion of the renormalization of the 
parameters in ${\cal L}_{ch}$.
Instead we concentrate on
an estimation of the size of the finite loop corrections
to the specific tree level terms shown in Fig.2.
This will be done by applying standard Feynman rules 
and using dimensional regularization \cite{bkm95}.
Specifically, we only need consider 
the loop corrections to the single $\pi^0 NN$ vertex 
in the s-wave channel:

\begin{equation}
%\Gamma_{\pi_0 NN} ^{s-wave}  
{\cal T}^{\mbox{\scriptsize Born}}_{-1}
+ {\cal T}^{\mbox{\scriptsize Corr}}_{+1} = 
\left ( {- \frac{g_A}{2 m_N f_\pi} } \right ) \;  
\sum_{i=1,2}
\left [ S_i \cdot (p_i' + p_i) \right ] \;
(v \cdot q) \;
\tau_i^0 \; \; 
{\cal V}
\label{Bornnew}
\end{equation}
where $S_i =  (0, \frac12 \bbox{\sigma}_i )$ is the spin of the
$i$-th proton and $\cal V$ is the amplitude to be calculated.
For the Born term [Fig.2(a)] itself we have:

\begin{equation}
{\cal V}_{2a}  = 1
\end{equation}
given by Eq.(\ref{Tm1}).
The loop diagrams [Figs.3(a)-(f)],
which renormalize the s-wave Born term, 
give the following contributions:

\begin{mathletters}
\label{loop}
\begin{equation}
{\cal V}_{3a} = 
\frac{1}{4} 
\left (
\frac{g_A}{f_\pi} 
\right ) ^2
\frac{J_2 (v p') - J_2 (v p) }{v q}
\label{loopa}
\end{equation}
\begin{equation}
{\cal V}_{3b} = 
- \frac{\Delta_\pi}{2 {f_\pi}^2}
\label{loopb}
\end{equation}
\begin{equation}
{\cal V}_{3c} = 
\frac{1}{2 {f_\pi}^2 } 
\frac{J_2 (v p') - J_2 (v p)}{v q}
\label{loopc}
\end{equation}
\begin{equation}
{\cal V}_{3d} = 
\frac{1}{2 {f_\pi}^2} 
\left (
3 \Delta_\pi + \left [ (v p) J_0 (v p) + (v p') J_0 (v p') \right ]
+ \frac{ (v p)^2 J_0 (v p) - (v p')^2 J_0 (v p') }{ v q } 
\right )
\label{loopd}
\end{equation}
\end{mathletters}
Here we have adopted the notations of Ref.\cite{bkm95}.
Thus
\begin{eqnarray}
\Delta_\pi &\equiv&\frac{1}{i} \int \frac{d^d l}{(2 \pi)^d}
\frac{1}{{m_\pi}^2  - l^2}\,=\,{m_\pi}^{d-2} (4 \pi)^{d/2}
\Gamma \left ( 1 - \frac{d}{2} \right )
\label{integral}\\
&=&2 {m_\pi}^2\left( 
L + \frac{1}{16 \pi^2 } \ln \frac{m_\pi}{\lambda}\right ),
\label{delta}
\end{eqnarray}
where the divergence is included in
\begin{equation}
L = \frac{\lambda^{d-4} }{16 \pi^2}
\left [ 
\frac{1}{d-4} +
\frac{1}{2} (\gamma_E - 1 - \ln 4\pi ) \right ].
\label{L}
\end{equation}
In this expression $\lambda$ denotes the dimensional 
regularization scale and $\gamma_E = 0.557215$.
Furthermore, $J_0$ and $J_2$ in Eqs.\ (\ref{loop}) are
defined by
\begin{equation}
J_0 (\omega) = 
- 4 L \omega 
+ \frac{\omega}{8 \pi^2} 
\left ( 1 - 2 \ln \frac{m_\pi}{\lambda} \right )
- \frac{1}{4 \pi^2}
\sqrt{{m_\pi}^2 - \omega^2 } 
\arccos{ \left ( \frac{-\omega}{m_\pi} \right )}
\label{j0}
\end{equation}
and
\begin{equation}
J_2 (\omega) = 
\frac{1}{d-1} \left [ ({m_\pi}^2 - \omega^2 ) J_0 (\omega) - 
\omega \Delta_\pi \right ]
\label{j2}
\end{equation}
The two contributions to $\cal V$, 
Eqs.(\ref{loopc}) and (\ref{loopd}),
originate from two different combinations of terms in Eqs.
(\ref{eq:17}).
To calculate Eq.(\ref{loopc}),
the second term in Eq.(\ref{eq:Hint0})  
and the first term in Eq.(\ref{eq:Hint1}) are used at the vertices,
whereas Eq.(\ref{loopd}) is calculated 
using the first term of Eq.(\ref{eq:Hint0})
and the second term of Eq.(\ref{eq:Hint1}).

The standard renormalization consists in the 
following procedure:\\
(1) The loop contributions to $\cal V$ are separated into 
a divergent part, which we take to be 
proportional to $L$ of Eq.(\ref{L}) and which contains
a pole at $d = 4$, and a finite part:
\begin{equation}
{\cal V}_{3} = 
{\cal V}_{3} \vert^{\infty} + 
{\cal V}_{3} \vert^{finite}
\end{equation}
(2) Local counter terms, which are of the 
same chiral order as the loop diagrams, 
are added.
In our case these counter terms must come from the Lagrangian 
${\cal L}^{(3)}$.
\begin{equation}
{\cal L}^{(3)} = \frac{1}{(4 \pi f_{\pi})^2} 
\sum_i D_i \bar{N} O_i N
\label{L3}
\end{equation}
to give two-nucleon diagrams with $\nu = 1$.
The unknown constants $D_i$ are then written as a sum of a finite 
and an infinite part
\begin{equation}
D_i = D_i \vert^{finite}(\lambda) 
+ ( 4 \pi)^2 \delta_i L. 
\end{equation}
The constants $\delta_i$ are determined
by requiring that 
the infinite part of $D_i$  cancel the divergent part
${\cal V}_{3} \vert^{\infty}$.
The remaining finite contributions which should be added
to the Born term via Eq.(\ref{Bornnew}), are 
\begin{equation}
{\cal V}_{loop} = 
{\cal V}_{3} \vert^{finite} +
{\cal V}_{c.t.} \vert^{finite}.
\end{equation}
The amplitude ${\cal V}_3$ contains energy-independent and 
energy-dependent parts,
as can be seen in Eq.\ (\ref{loop}).
The energy-independent part can be absorbed 
in the renormalization of 
the following physical parameters:
the pion wave function renormalization factor $Z_\pi$ [Fig.3(e)],
the nucleon mass $m_N$ and the nucleon wave function renormalization
factor $Z_N$ [(Fig.3(f)], 
as well as
the axial coupling constant $g_A$ [Figs.3(a),(b),(e),(f)].
For the evaluation of the energy-dependent part
we use the {\it typical threshold kinematics}:
$v q = m_\pi$, $v p_1 = m_\pi$, $v p_2 =0$.
Putting these values into the corresponding terms 
in eq.\ (\ref{loop}),
we obtain as the total contribution 
of the diagrams in Fig.3
\begin{equation}
{\cal V}_{3} \vert^{finite} \approx 0.1.
\end{equation}
Thus ${\cal V}_{3} \vert^{finite}$ amounts to
$10\%$ of the Born term [Fig.2(a)].
In addition we have finite contributions from the counterterms of
${\cal L}^{(3)}$, ${\cal V}_{c.t.} \vert^{finite}$. 
We note that only very few 
of the low energy constants in the counterterms 
$D_i \vert^{finite} (\lambda)$ are known \cite{bkm95}.
Some of the low energy constants in ${\cal L}^{(2)}$,
$B_i \vert^{finite} (\lambda)$ have been
estimated in Ref.\cite{bkm95} 
assuming $\Delta$ resonance saturation.
The result indicates
$B_i \vert^{finite} (\lambda) \approx {\cal O} (0.1)$. 
For an estimate of the low energy constants 
$D_i \vert^{finite} (\lambda)$
in ${\cal L}^{(3)}$
it seems reasonable to assume
that they
are of the same order of magnitude 
as the $B_i \vert^{finite} (\lambda)$ in ${\cal L}^{(2)}$.
To be conservative let us assume 
$D_i \vert^{finite} (\lambda) \approx {\cal O} (1)$;  
then we expect
${\cal V}_{c.t.} \vert^{finite} \approx 0.1$.
It is clear that, if those coefficients were ``unreasonably large'', 
the convergence of the whole chiral series 
would be destroyed.

Altogether, after renormalization 
the total contributions from the loop terms 
are expected to amount to at most $20 \%$ 
of the Born term. 
This is not a completely negligible contribution
in the present context because,
as will be discussed in the next section,
there can be a significant cancelation
between the Born and the rescattering terms.
Nevertheless, since our present treatment
involves other larger uncertainties,
we will neglect the renormalization
of the Born term and henceforth concentrate 
on the bare Born term [Fig.2(a)]
and the rescattering term [Fig.2(b)].

\section{Calculation of the Two-Nucleon Transition Matrix}
We derived in section IV 
the effective transition operator ${\cal T}$
arising from the tree diagrams
and, in section V, we estimated 
the additional contributions due to the loop corrections 
and presented an argument for ignoring the loop corrections
in this work.
These considerations lead to 
the the HFF expression of ${\cal T}$ 
up to order $\nu$ = 1, 
given in Eqs.(\ref{eq:calTtrunc}) and (\ref{eq:35}),
and this ${\cal T}$ is to be used in Eq.(\ref{eq:Tmatrix})
to obtain the two-nucleon transition matrix $T$.

A formally ``consistent" treatment of 
Eq.(\ref{eq:Tmatrix})
would consist in using for $|\Phi_i\rangle$ 
and $|\Phi_f\rangle$ 
two-nucleon wave functions 
generated by irreducible diagrams
of order up to $\nu=1$. 
A problem in this ``consistent" $\chi$PT approach is
that the intermediate two-nucleon propagators 
in Fig.1 can be significantly off-mass-shell,
which creates a difficulty in any $\chi$PT calculation.
Another more practical problem is 
that, if we include the initial- and final- 
two-nucleon ($N$-$N$) interactions
in diagrams up to chiral order $\nu=1$, 
these $N$-$N$ interactions are not 
realistic enough to reproduce the known $N$-$N$ observables.
A pragmatic remedy for these problems is to use 
a phenomenological $N$-$N$ potential
to generate the distorted $N$-$N$ wavefunctions. 
Park, Min and Rho \cite{pmr95} used 
this hybrid approach to study the exchange-current
in the $n+p \rightarrow \gamma+d$ reaction and 
at least, for the low-momentum transfer process
studied in Ref.\cite{pmr95},
the hybrid method is known to work extremely well.

Apart from the above-mentioned problem,
there is a delicate aspect in the derivation 
of an effective two-body operator
from a given Feynman diagram.
Ordinarily, one works with ${\bf r}$-space 
transition operators
acting on ${\bf r}$-representation wavefunctions,
for the nuclear wavefunctions are commonly given 
in this representation.
To this end, a Feynman amplitude 
which is most conveniently given 
in momentum space, is Fourier-transformed
into the ${\bf r}$-representation.
This method works best 
for low momentum transfer processes
which have substantial transition amplitudes
for on-shell initial and final plane-wave states.
However, the $pp \rightarrow pp\pi^0$ reaction at threshold
does not belong to this category.
For this reaction it is essential to recognize 
that the nucleon lines that appear as external lines
in Fig.2 are in fact internal lines 
in larger diagrams illustrated in Fig.1.
These internal lines can be far off-shell
due to the initial- and final-state interactions.
Indeed without this off-shell kinematics,
the Born term [Fig.2(a)]
would not contribute at all !
In the conventional approach, however,
one ignores this feature in deriving
${\cal T}$ in coordinate representation.
For example, in Fourier-transforming 
an operator of the type of
${\cal T}^{\mbox{\scriptsize Res}}_{+1}$,
Eq.(\ref{Tp1}),
even though $\bbox{p}_i$ and $\bbox{p^\prime}_i$ 
in Eq.(\ref{Tp1}) in fact can be anything 
due to momentum transfers
caused by the initial and final $N$-$N$ interactions,
it is a common practice to keep the energy
of the propagating pion fixed 
at the value determined by the asymptotic energies 
of the nucleons. 
Hanhart et al.\cite{hanetal95}
made a critical study of 
the consequences of avoiding these 
kinematical approximations. 
They worked directly with the 
two-nucleon wavefunctions
in momentum representation.
In the present work we do not attempt at detailed 
momentum-space calculations and
simply use the ``conventional" Fourier transform method.
Because of this and a few other approximations
adopted, the numerical work presented here 
is admittedly of exploratory nature.
Nonetheless, as we shall show,
our semiquantitative study of $T$
based on the chiral-theoretically motived
transition operator ${\cal T}$
provide some valuable insight into 
the dynamics of the threshold
$pp \rightarrow pp\pi^0$ reaction.

Let us denote the contribution of Fig.2(b)
for plane-wave initial and final states by
$<\!p_1',\,p_2',\,  q|{\cal T}^{(1)}|p_1,\,p_2\!>$.
We first calculate this matrix element 
for the {\it typical threshold kinematics} described earlier;
for the meson variables,
$q \!=\! (m_\pi,\bbox{0})$ and
$k \!=\! (m_\pi/2,\bbox{k})$ 
with $|{\bf k}|\,=\,\sqrt{m_\pi m_N}$.
Correspondingly, the coupling strength
$\kappa(k,q)$ [Eq.(\ref{Tp1})]
is taken to be $\kappa_{th}\,=\,- 1.5$GeV$^{-1}$
[Eq.(\ref{kappath1})].
Subsequently, by liberating the momentum variables
${\bf p}_1$, ${\bf p}_1'$, 
${\bf p}_2$, and ${\bf p}_2'$ 
from the on-mass-shell conditions ($p_1^2=m_N^2,\ldots$),
we treat $<\!p_1',\,p_2',\,q|{\cal T}^{(1)}|p_1,\,p_2\!>$
as a function of ${\bf p}_1$, ${\bf p}_1'$, 
${\bf p}_2$, and ${\bf p}_2'$.
Let $T({\bf p_1'},\,{\bf p_2'};\,{\bf p_1},\,{\bf p_2})$
stand for this function.
We still require momentum conservation at each vertex, 
which imposes the conditions 
${\bf p}_1+{\bf p}_2\,=\,
{\bf p}_1'+{\bf p}_2'+{\bf q}=0$,
and ${\bf k}={\bf p}_1' - {\bf p}_1 
={\bf p}_2 - {\bf p}_2'$.
$T({\bf p_1'},\,{\bf p_2'};\,{\bf p_1},\,{\bf p_2})$
can be easily Fourier transformed to give
$\tilde{\cal T}^{\mbox{\scriptsize Res}}_{+1}$
in ${\bf r}$ representation.
The simplified treatment described here,
which is commonly used in the literature, 
shall be referred to as the
{\it fixed kinematics approximation}.

Now, in the {\it fixed kinematics approximation},
${\cal T}$ [Eqs.(\ref{eq:calTtrunc}), (\ref{eq:35})]
is translated into differential operators acting 
on relative coordinate of the
two-nucleon wavefunctions:

\begin{mathletters}
\label{eq:37}
\begin{equation}
\tilde{\cal T}^{\mbox{\scriptsize Born}}_{-1}
 = \frac{g_A}{f_\pi} \frac{m_\pi}{m_N} 
  \bbox{\Sigma} \!\cdot\! \bbox{\nabla}_r, 
\label{eq:Tp-1r}
\end{equation}
\begin{equation}
\tilde{\cal T}^{\mbox{\scriptsize Res}}_{+1}
 = - \frac{2g_A}{f_\pi} \kappa_{th} 
\bbox{\Sigma}\! \cdot\! \hat{\bbox{r}} f^\prime(r),
\label{Tp1r}
\end{equation}
\end{mathletters}
where the derivative operator with subscript $r$ 
is to act on the relative coordinate 
$\bbox{r}$ between two protons,
and 
$\bbox{\Sigma} \equiv \frac12 (\bbox{\sigma}_1 - 
\bbox{\sigma}_2)$. 
The trivial isospin operator $\tau^0_i$ 
has been dropped.
The Yukawa function 
$f(r) \equiv \exp(-\mu^\prime r)/4\pi r$
is defined with the {\it effective} mass 
$\mu^\prime = \frac{\sqrt3}{2} m_\pi$.   
We reemphasize that 
the simple Yukawa form $f(r)$ arises 
only when the {\it fixed kinematics approximation} 
just discussed is used.

From this point on, our calculation of
$T$ follows exactly the traditional pattern 
described in the literature.
Thus $T$ is evaluated 
by inserting the transition operators,
$\tilde{\cal T}^{\mbox{\scriptsize Born}}_{-1}$
and 
$\tilde{\cal T}^{\mbox{\scriptsize Res}}_{+1}$,
Eq.(\ref{eq:37}), 
between the initial and final nuclear states 
\begin{equation} 
\begin{array}{l}
\phi_i(\bbox{r}) = (\sqrt2/pr) 
iu_{1,0}(r) e^{i\delta_{1,0}} (4\pi)^{1/2} 
   |{}_{}^{3,3}P^{}_0\rangle, \\
\phi_f(\bbox{r}) = 
(1/p^\prime r) u_{0,0}(r) 
e^{i\delta_{0,0}} (4\pi)^{1/2}
   |{}_{}^{3,1}S^{}_0\rangle, 
\end{array} 
\end{equation}
where $p$ and $p^\prime$ are 
the asymptotic relative three-momenta 
of the initial and final two-proton systems. 
The wavefunctions are normalized as 
$u_{L,J} \stackrel{r \rightarrow \infty}
\longrightarrow 
\sin (pr - \frac12\pi L + \delta_{L,J})$ 
with $\delta_{L,J}$ 
being the $N$-$N$ scattering phase shifts.
For simplicity, 
the Coulomb interactions 
between the two protons is ignored.  
(The Coulomb force is known 
to reduce the cross section 
up to 30\%.\cite{ms91}.) 
The explicit expression for the
transition amplitude at threshold is obtained as 
\begin{equation}
T(E_f) = 
4 \pi (g_A/f_\pi m_N m_\pi^{3/2}) 
(J_{-1}^{\mbox{\scriptsize Born}} + 
J_{+1}^{\mbox{\scriptsize Res}}), 
\end{equation}
Here, $E_f = E_{p^\prime} + \bbox{q}^2/2m_\pi$ 
is the kinetic energy of the final state, and 
\begin{mathletters}
\label{eq:40}
\begin{equation}
 J_{-1}^{\mbox{\scriptsize Born}} = 
\lim_{p^\prime\rightarrow 0}
\frac{-m_\pi^2}{pp^\prime} 
\int^\infty_0 \!\! dr r^2 \frac{u_{0,0}}{r} 
\left( \frac{d}{dr} + \frac{2}{r} \right) 
\frac{u_{1,0}}{r}, 
\label{eq40a}
\end{equation}
\begin{equation}
 J_{+1}^{\mbox{\scriptsize Res}} = 
\lim_{p^\prime\rightarrow 0}
2\kappa_{th} \frac{m_\pi M_n}{pp^\prime} 
\int^\infty_0 \!\! dr u_{0,0} f^\prime(r) u_{1,0}.
\label{eq40b}
\end{equation}
\end{mathletters}
The total cross section is obtained 
by multiplying the 
absolute square of the transition amplitude
(averaged over the initial spins
and summed over the final spins)
with the appropriate phase space factor $\rho(E_f)$
and the flux factor $1/v$:
\begin{equation}
\sigma_{\mbox{\scriptsize\it tot}}^{} = 
\frac{2\pi}{v} \int \!\! d\rho(E_f) 
|T(E_f)|^2. 
\end{equation}
For a rough estimation
one may approximate the energy dependence of 
the transition matrix as\cite{wat52} 

\begin{equation}
|T(E_f)|^2 = \frac{|T(0)|^2}{1 + p^{\prime 2}a^2},
\label{aprx}
\end{equation}
where $a$ is the scattering length 
of the $NN$ potential. 
Then the cross section can be 
simply expressed as 

\begin{equation}
\sigma_{\mbox{\scriptsize\it tot}} 
= \frac{g_A^2}{\sqrt2\pi f_\pi^2 m_\pi^2} 
|J|^2 I(E_f)
\label{cs}
\end{equation}
where

\begin{mathletters}
\label{eq:44}
\begin{equation}
|J|^2 = |J_{-1}^{\mbox{\scriptsize Born}} + 
J_{+1}^{\mbox{\scriptsize Res}}|^2 ,
\label{eq44a}
\end{equation}
\begin{equation}
I(E_f) = \int^{E_f}_0 \!\!dE_{p^\prime} 
\frac{\sqrt{E_f - E_{p^\prime}}
\sqrt{E_{p^\prime}}}{1 + m_N a^2 E_{p^\prime}} .
\label{eq44b}
\end{equation}
\end{mathletters}
Under the approximation (\ref{aprx}), 
the energy dependence 
of the cross section is solely given by $I(E_f)$,
which incorporates the phase space
and the final state interaction effect 
(in the Watson approximation\cite{wat52}).

We have calculated 
the integrals $J_{-1}^{\mbox{\scriptsize Born}}$
and $J_{+1}^{\mbox{\scriptsize Res}}$ 
for representative nuclear potentials:
the Hamada-Johnston (HJ) potential\cite{hj62},
and the Reid soft-core potential (RSC)\cite{rei68}.
The results are given in Table I,
and the corresponding cross sections 
are presented in Table II. 
These results indicate that,
for the nuclear potentials considered here,
the value of $|J|$ is much too small 
to reproduce the experimental cross section. 
If we define the discrepancy ratio $R$ by
\begin{equation} 
R \equiv \sigma_{tot}^{exp}/\sigma_{tot}^{calc} \, ,
\label{ratio}
\end{equation}
with $\sigma_{tot}^{exp}$ taken from Ref.\cite{meyetal90},
then $R\,\cong\,80$ ($R\,\cong\,210$)
for the Hamada-Johnston (Reid soft-core) potential, 
and $R$ happens to be almost constant
for the whole range of $E_f\leq 23$ MeV for which
$\sigma_{tot}^{exp}$ is known.
Thus, although the off-shell behavior 
of the $s$-wave pion scattering amplitude 
derived from the chiral Lagrangian 
does enhance the contribution 
of the rescattering process 
over the value reported in the literature, 
the sign change that occurs in $\kappa$ 
as one goes from $\kappa_0$ 
[Eq.(\ref{eq:onshell})] to $\kappa_{th}$ 
[Eq.(\ref{eq:kappath})]
results in a significant cancelation 
between the Born term 
$J_{-1}^{\mbox{\scriptsize Born}}$ 
and the rescattering term
$ J_{+1}^{\mbox{\scriptsize Res}}$,
leading to the very small cross sections
in Table II \cite{cfmv95a}.
The drastic cancelation 
between $J_{-1}^{\mbox{\scriptsize Born}}$
and $ J_{+1}^{\mbox{\scriptsize Res}}$
found here means that the calculated cross sections
are highly sensitive to the various approximations
used in our calculation
and also to the precise values of the constants 
$c_1, c_2$ and $c_3$ of Eq.(\ref{eq:20}).
We will discuss these two questions in the next two paragraphs.

We adopted the threshold kinematics approximation
and neglected the energy-momentum 
dependence in Eq.(\ref{eq:kappakq}) and
treated the vertices in Figs.1 and 2 
as fixed numbers, {\it i.e.}, 
$\kappa(k,q)$ = $\kappa_{th}$ = constant.
In addition, although 
the loop corrections of chiral order $\nu$ = 1, shown in Fig.3(a),
automatically introduces energy-momentum dependent
vertices, we ignored this feature. 
The fact that the kinematics of the reaction 
Eq.(\ref{eq:pppi}) requires highly off-shell vertices
leads to the expectation 
that the vertex form factors can be very important
and invalidate the 
{\it threshold kinematics approximation}
leading to Eq.(\ref{eq:37}). 
In this connection we note that 
a momentum-space calculation \cite{hanetal95},
which is free from this approximation,
indicates that even a negative value of $\lambda_1$ 
could lead to the moderate enhancement of the cross section.  

The strong cancelation between 
the Born and rescattering terms
also means that, even within the framework of
the {\it fixed kinematics approximation},
the large errors that exist in the empirical value of 
$a^+$ and the $c_1, c_2$ and $c_3$ constants
can influence the cross sections significantly.
To assess this influence, we rewrite 
Eq.(\ref{eq:kappath}) as
\begin{equation}
\kappa_{th}\,=\,\frac{m_\pi^2}{f_\pi^2}c_1
-\pi (1 + \frac{m_\pi}{m_N}) a^+
+ \frac{3g_A^2 m_\pi^3}{256\pi f_\pi^4}.
\end{equation}
The use of the experimental values 
for $a^+$ and $c_1$ quoted earlier leads to
\begin{equation}
\kappa_{th}\,=\,(-1.5 \pm 0.4)\,{\rm GeV}^{-1}.
\end{equation}
With this uncertainty taken into account,
the ratio $R$ ranges from 
$R=25$ to $R=2100$ for the Hamada-Johnston potential,
and from $R=50$ to $R=3.4\times 10^4$ 
for the Reid soft-core potential.
To further examine the uncertainties in the ${\cal L}^{(1)}$ constants 
we remark that the
value of $c_2 + c_3$ can be extracted from the known pion-nucleon
effective range parameter $b^+$.  
The low energy pion-nucleon scattering amplitude is expanded
as:
\begin{equation}
f^+ = a^+ + b^+ \bbox{q}^2 + \cdots
\end{equation}
where {\bf q} is the pion momentum and 
$b^+ = (-0.044 \pm 0.007) m_{\pi}^{-3}$ \cite{hoe83}.
If we use ${\cal L}^{(1)}$ to calculate the s-wave 
pion-nucleon amplitude we find:
\begin{equation}
b^+ = \frac{1}{2 \pi} \left( 1 + \frac{m_{\pi}}{m_N} \right)^{-1}
\left( \frac{m_{\pi}}{f_\pi} \right)^2 ( c_2 + c_3 - \frac{g_A^2}{8 m_N})
\frac{1}{{m_\pi}^2} \, ,
\label{bplus}
\end{equation}
and then Eq.(\ref{eq:kappath}) leads to
\begin{eqnarray}
\kappa_{th} &= \frac{2 {m_\pi}^2 }{{f_\pi}^2} c_1 - \pi {m_\pi}^2
\left ( 1 + \frac{m_\pi}{m_N} \right ) b^+
\nonumber \\
&= (-2.7 \pm 0.6) \mbox{GeV}^{-1} 
\label{kappanew}
\end{eqnarray}
Since $c_3$ is given directly by the experimental quantity $\alpha_A$ 
[Eq.(\ref{C2})], we consider Eq.(\ref{bplus}) as an alternative
input to determine $c_2$ in terms of $b^+$ and $c_3$.
Then Eqs.(\ref{bplus}),(\ref{C2}) and the experimental value 
of $b^+$ \cite{hoe83} give
\begin{equation}
c_2 = (4.5 \pm 0.7 ) {\rm GeV}^{-1} \, .
\label{c2new}
\end{equation}
We note that this value is larger than the one given
in Eq.(\ref{C3}), indicating that the determination of $c_2$ requires
further studies.
With the new value of $\kappa_{th}$ given in Eq.(\ref{kappanew})
we find that the discrepancy ratio $R$ [Eq.(\ref{ratio})] can be
as small as $\sim 10$.
( In this case 
$\vert J_{+1}^{\mbox{\scriptsize Res}} \vert > 
\vert J_{-1}^{\mbox{\scriptsize Born}} \vert$; 
the exact
cancelation between the Born and the pion rescattering term
occurs for $\kappa_{th} \sim - 2 {\mbox{GeV}}^{-1}$.) 

Without attaching any significance
to the detailed numbers above,
we still learn  
the extreme sensitivity of $\sigma_{tot}^{calc}$
to the input parameters
and that, despite this high sensitivity,
$\sigma_{tot}^{calc}$ still falls far short of 
$\sigma_{tot}^{exp}$ (within the framework of
the {\it fixed kinematics approximation}).

%We should also note 
%the strong dependence of the integrals, 
%$J^{\mbox{\scriptsize Born}}_{-1}$
%and $J^{\mbox{\scriptsize Res}}_{+1}$,
%on the two different $N$-$N$ potentials.
%This dependence might be ascribed to the fact that, 
%although all the potentials used here
%yield more or less the same phase shifts,
%they give very different off-shell behaviors, 
%especially for the ${}^1 S_0$ state \cite{mac89}. 
%The $pp \rightarrow pp\gamma$ process 
%has been studied with the view to 
%obtaining information on the off-shell structure 
%of the nucleon potentials {\it REALLY!!???}. 
%However, this process is mainly 
%described by the $^3 P_0$ state 
%whose off-shell structure exhibits
%rather weak dependence on nuclear interactions. 

\section{Discussion and Conclusions}

In this work we have used $\chi$PT 
to calculate the effective pion-exchange current contribution
to the $ p p \rightarrow p p \pi^0$ reaction at threshold. 
As stated repeatedly,  
our aim here is to carry out 
a systematic treatment of ${\cal T}$
up to chiral order $\nu=1$ [see Eq.(\ref{eq:calTtrunc})].
However, in order to make contact 
with the expressions appearing in the literature  
\cite{kr66}, let us consider 
a very limited number of $\nu=2$ diagrams.
To be specific, we consider a diagram
in Fig.2(b) but with the ${\bar \nu}=0$ 
($p$-wave) $\pi NN$ vertex
replaced with a $\bar{\nu}=1$ ($s$-wave) vertex.
Then, instead of Eq.(\ref{Tp1r}), we will obtain 
\begin{equation}
{\cal T}^{\mbox{\scriptsize Res}}_{1+2} 
= -\frac{g_A}{f_\pi} \kappa_{th} 
\bbox{\Sigma} \!\cdot\! 
\left[ \hat{\bbox{r}} f^\prime(r) 
\left( 2 \!+\! \frac{m_\pi}{2m_N} \right) 
\!+\! f(r)\frac{m_\pi}{m_N} \bbox{\nabla}_r 
\right],  
\label{Textra}
\end{equation}
which is the two-body transition operator
used in Ref.\cite{kr66}.
Thus, 
we do recover the usual 
phenomenological parameterization in $\chi$PT,
but this is just one of many $\nu =2$ diagrams. 
Our systematic $\nu =1$ calculation excludes all $\nu =2$ diagrams.

We have also ignored the 
exchange current contributions from scalar and
vector two-nucleon exchanges. 
Following the $\chi$PT of Refs. \cite{wei90,kol92} 
the vector meson
exchange is largely accounted 
for via the four-nucleon contact terms
illustrated in Fig.4(a).
If we had retained the last two terms of Eq.(\ref{eq:11}),
the pion-nucleon interaction ${\cal H}^{(1)}_{int}$,
Eq.(\ref{eq:Hint1}),
would have had an additional piece 
${\cal H}^{(1)'}_{int}$

\begin{eqnarray}
{\cal H}^{(1)'}_{int} &=&
\frac{c_9}{4m_N f_\pi} (\bar{N}N)
(\bar{N} \bbox{\sigma} \!\cdot\!\bbox{\nabla}
(\bbox{\tau} \!\cdot\!\bbox{\pi}) N) 
\nonumber\\
&& 
+ \frac{c_{10}}{4m_N f_\pi} 
(\bar{N} \bbox{\sigma} N) \cdot
(\bar{N} \bbox{\nabla} (\bbox{\tau} 
\!\cdot\!\bbox{\pi}) N). 
\label{eq:Hcontact}
\end{eqnarray}
The ${\cal H}^{(1)'}_{int}$ term of 
Fig.4(a) has a 
$\bbox{\sigma} \!\cdot\! \bbox{q}$ structure,
which means it describes $p$-wave pion production
and therefore does not contribute 
to the threshold $pp \rightarrow pp\pi^0$ reaction.
The $s$-wave pion production contact term, 
also belonging to the type of diagram illustrated in Fig.4(a), 
enters as a $\frac{1}{m_N}$ recoil correction 
to ${\cal H}^{(1)'}_{int}$
and therefore is of chiral order $\nu$ = 2.
Formally, the chiral order $\nu=2$ 
diagrams have no place
in the present calculation limited to $\nu=1$.
However, in view of the great current interest
in the possible large contribution of
the heavy-meson exchange diagrams,
we make a few remarks 
on the s-wave $\nu=2$ contact terms
depicted in Fig.4(a). 
We note that the coordinate representation 
of this contact term contains $\delta^3(\bbox{r})$.
Meanwhile, 
in the threshold $pp \rightarrow pp\pi^0$ reaction 
the initial two-nucleon relative motion
must be in $p$-wave (because of parity)
and so its wavefunction
vanishes at $\bbox{r}=0$. 
Thus, even in a chiral order $\nu=2$ calculation, 
the contact term Fig.4(a) corresponding to 
s-wave pion production will play no role.
Including meson loops corrections
to these contact terms [an example illustrated in Fig.4(b)] 
would smear out the $\delta$-function behavior,
allowing them to have a finite contribution 
to the threshold $pp \rightarrow pp\pi^0$ reaction.
This involves, however, diagrams 
of even higher chiral order than $\nu=2$. 
Thus, in order to include the strong effective 
isoscalar-vector repulsion of the $N$-$N$ forces
($\omega$ exchange) contained 
in the four-nucleon contact terms of Weinberg's \cite{wei90} 
and van Kolck's {\it et al.}'s \cite{kol92} 
$\chi$PT description, we have to
go to chiral order $\nu = 3$.

Meanwhile, one may picture the ``effective heavy mesons" 
as generated by multi-pion exchange diagrams like those 
illustrated in Fig.5.
These diagrams, which necessarily contain loops,
represent a very limited class of $\nu \ge 3$ diagrams.
For example, an important  
part of the effective scalar exchange
between two nucleons involve 
intermediate $\pi$-$\pi$ $s$-wave interaction 
which requires at least two loop diagrams like Fig.5(c).
Thus, if we are to interpret 
the heavy-meson exchange diagrams 
of Lee and Riska\cite{lr93}
in the framework of 
nuclear $\chi$PT,
we must deal with terms with 
chiral order $\nu \geq 3$,
which at present is beyond practical calculations. 

We now recapitulate 
the main points of this article.
\begin{enumerate} 
\item
Using $\chi$PT in a systematic fashion 
we have shown that the contribution of 
the pion rescattering term 
can be much larger than obtained in the traditional 
phenomenological calculations. 
This fact itself 
supports the suggestion of Hernandez and Oset \cite{ho95}
that the off-shell $s$-wave pion-nucleon scattering
should enhance the rescattering contribution significantly.
However, the sign of the enhanced rescattering vertex
obtained in $\chi$PT is {\it opposite}
to that used in Ref.\cite{ho95},
at least for the {\it typical threshold kinematics} 
defined in the text.
This sign change in the coupling constant 
$\kappa_{th}$ 
leads to a destructive interference 
between the Born and rescattering terms
instead of the constructive interference found 
in Ref.\cite{ho95}. 
The significant cancelation between these terms
give rise to the very small cross section
for the near-threshold 
$pp \rightarrow pp\pi^0$ reaction 
calculated in this work.
Although our particular numerical results were 
obtained in what we call the {\it fixed kinematics 
approximation}, 
these results at least indicate 
that the large enhancement of $\sigma_{tot}^{calc}$
obtained in Ref.\cite{ho95} is open to more
detailed examinations.

\item
The {\it fixed kinematics approximation} 
(which is commonly used in the literature)
should be avoided. There are 
at least two reasons why this is not a good
approximation for this reaction:
(i) The initial- and final-state interactions
play an essential role in the near-threshold 
$pp \rightarrow pp\pi^0$ reaction;
(ii) The theoretical cross section
within the framework of the Born plus 
rescattering terms is likely to depend on
the delicate cancelation between these two terms.
In a momentum space calculation \cite{hanetal95},
we can easily avoid the {\it fixed kinematics approximation}.
Such a calculation will allow us to work
with full off-shell kinematics, 
to incorporate the $\chi$PT form factors
in the Born term, and to reduce ambiguities 
in our calculation down to the level of uncertainties 
in the input parameters in $\chi$PT 
and the chiral counter terms.

\item
Several works \cite{lr93,hmg94,hanetal95} indicate 
that the two-nucleon scalar (sigma) exchange 
can be very important.
We gave in the introduction 
a simple kinematical argument for its plausibility,
and our dynamical calculation 
(albeit of semiquantitative nature)
seems to indicate the necessity of 
the sigma exchange contribution
in order to explain the observed cross sections
for the threshold $pp \rightarrow pp\pi^0$ reaction.
It is of great interest to see to what extent 
an improved $\chi$PT calculation based on
momentum-space representation helps 
sharpen the conclusion on the necessity of
the sigma exchange diagram.
Such a calculation is now in progress.
If it is established that 
the heavy meson exchange diagrams play an essential role in
the threshold $pp \rightarrow pp\pi^0$ reaction,
it seems that we must resort to a modified version of $\chi$PT,
for a brute force extension of our treatment to $\nu \geq 2$ seems 
extremely difficult.
An attempt to include vector meson degrees of freedoms 
explicitly can be found e.g. in Ref.\cite{pmr93}.
A purely phenomenological approach as used in \cite{lr93} 
may also be a useful alternative.

\end{enumerate}

\acknowledgements

We are grateful to U. van Kolck for the
useful communication on Ref. \cite{cfmv95}.
One of us (B.-Y. P) is grateful 
for the hospitality of 
the Nuclear Theory Group 
of the University of South Carolina,
where the main part of this work was done. 
This work is supported in part 
by the National Science Foundation, 
Grant No. PHYS-9310124.

%%%%%%%%%%%
%%%%%%%%%%%   References 
%%%%%%%%%%%

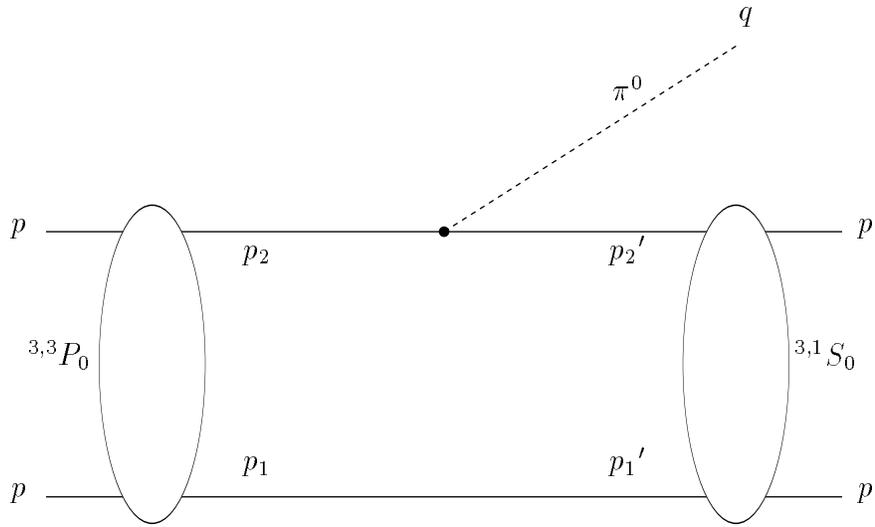
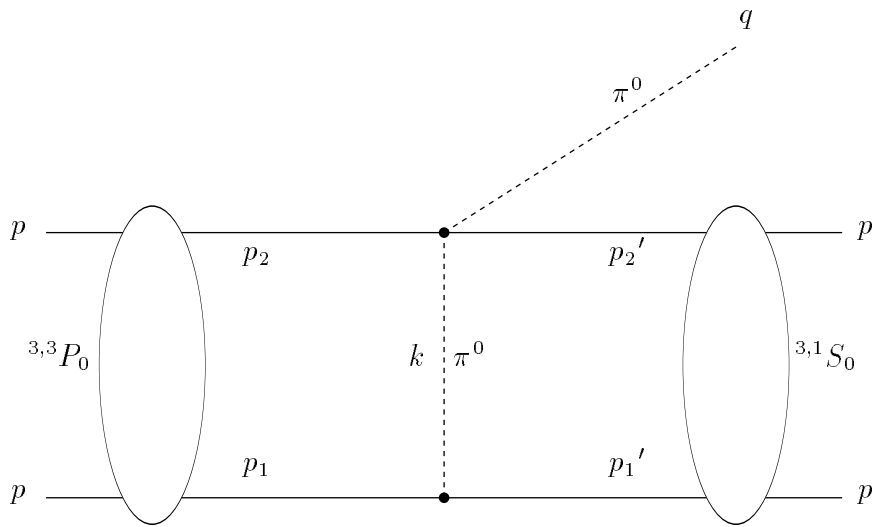
\begin{figure}
\caption{Single nucleon process (Born term) [Fig.1(a)] and
pion rescattering process for the $pp \to pp\pi^0$ reaction near
threshold. $^{2T+1,2S+1}L_J $ 
denotes the isospin and angular momenta
of the initial and final states.}
\end{figure}
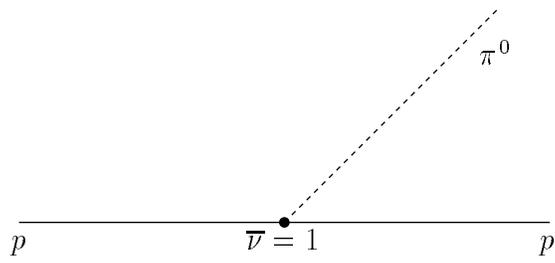
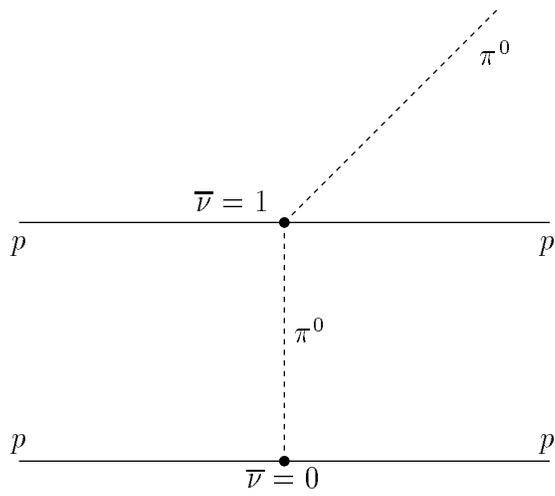
\begin{figure}
\caption{Tree graphs: The Born term [Fig.2(a)] ($\nu = -1$)
and the pion rescattering term [Fig.2(b)] ($\nu = 1$).}
\end{figure}
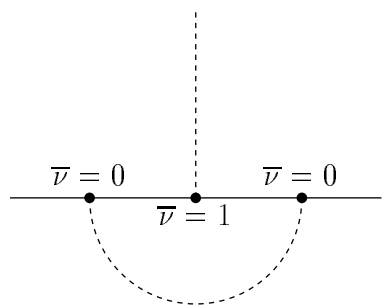
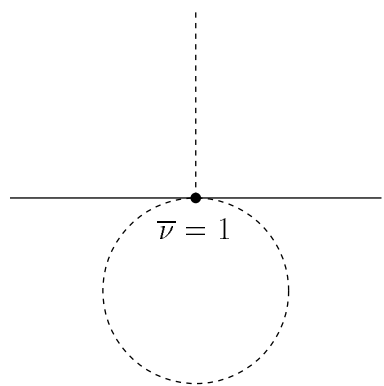
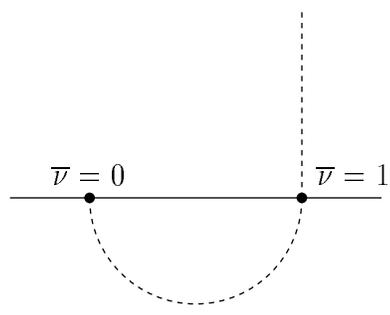
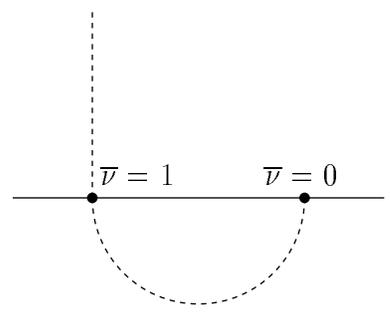
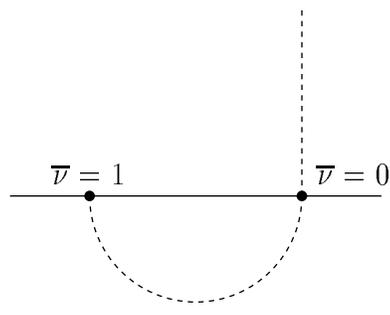
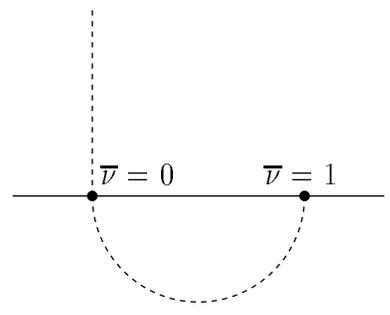
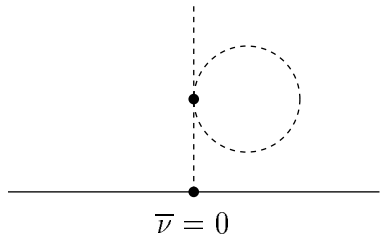
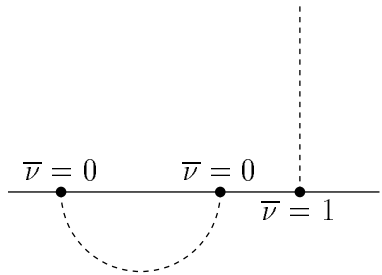
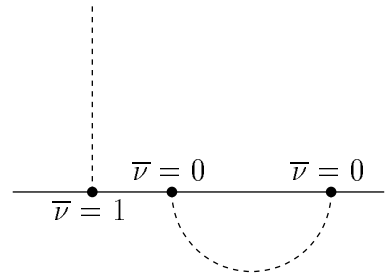
\begin{figure}
\caption{Loop corrections to the Born term.}  
%($\nu = 1$).}
\end{figure}
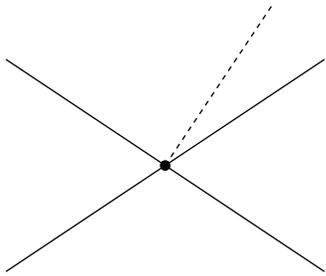
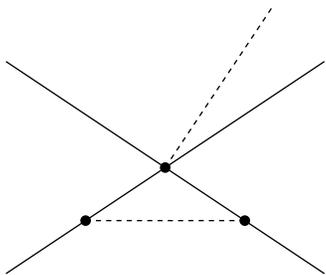
\begin{figure}
\caption{Generic four-fermion-pion vertex (contact term)
[Fig.4(a)] and an example of a 
loop correction to a contact term [Fig.4(b)].}
\end{figure}
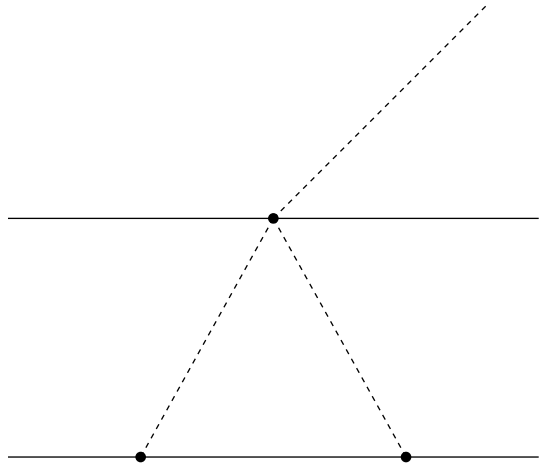
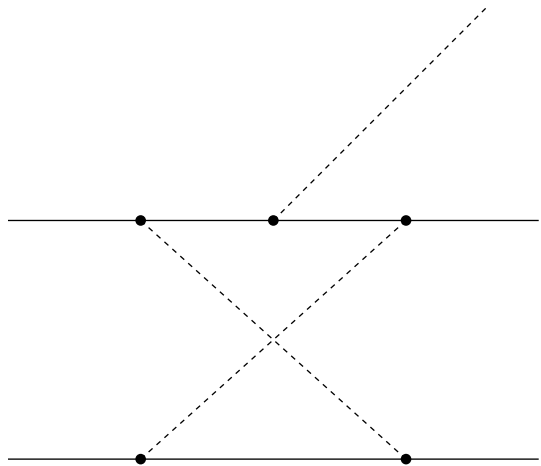
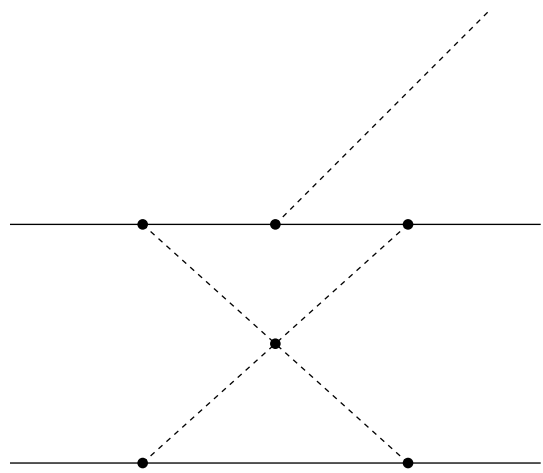
\begin{figure}
\caption{A few higher order diagrams 
contributing to the effective 
%$\sigma$
two-nucleon scalar exchange in nuclear $\chi$PT.}
\end{figure}

\begin{table} %---------------- Table 1 ---------------
\caption{
$J_{-1}^{\mbox{ Born}}$
and 
$J_{+1}^{\mbox{ Res}}$
for the threshold kinematics
[Eqs.(\protect{\ref{eq40a}}),(\protect{\ref{eq40b}})],
calculated with the Hamada-Johnsotn (HJ) 
and Reid soft-core (RSC) potentials.}
\begin{tabular}{ccc}
  & HJ & RSC \\
\hline
$J_{-1}^{\mbox{\scriptsize Born}}$ 
& $-0.672$ & $-0.515$  \\
$J_{+1}^{\mbox{\scriptsize Res}}$ 
& $+0.505$ & $+0.413$ \\
\end{tabular} 
\end{table} %----------------- Table 1 ----------------

\begin{table} 
%---------------- Table 2 ---------------
\caption{The total cross sections (in $\mu b$) 
as functions of 
$ \eta \equiv \protect{\sqrt{ 2{E_f} /  {m_\pi} }  }$,
calculated with the
Hamada-Johnston (HJ) and Reid soft-core (RSC) potentials.} 
\begin{tabular}{ccc}
$\eta$  & $\sigma_{HJ}$ & $\sigma_{RSC}$ \\
\hline
  .03     &     .0000  &   .0000  \\   
  .06     &     .0003  &   .0001  \\   
  .09     &     .0011  &   .0004  \\   
  .12     &     .0024  &   .0009  \\   
  .15     &     .0043  &   .0016  \\   
  .18     &     .0069  &   .0026  \\   
  .21     &     .0100  &   .0037  \\   
  .24     &     .0138  &   .0052  \\ 
  .27     &     .0182  &   .0068  \\ 
  .30     &     .0232  &   .0087  \\ 
  .33     &     .0289  &   .0108  \\ 
  .36     &     .0352  &   .0131  \\ 
  .39     &     .0421  &   .0157  \\ 
  .42     &     .0496  &   .0185  \\ 
  .45     &     .0577  &   .0215  \\ 
  .48     &     .0665  &   .0248  \\ 
  .51     &     .0759  &   .0283  \\ 
  .54     &     .0859  &   .0320  \\ 
  .57     &     .0965  &   .0360  \\ 
  .60     &     .1078  &   .0402  \\ 
\end{tabular} 
\end{table} 


\begin{references}

\bibitem{meyetal90}
H. O. Meyer {\it et al.\/}, 
Phys. Rev. Lett. {\bf 65}, 2846 (1990);
Nucl. Phys. {\bf A539}, 633 (1992).

\bibitem{Uppsala} A. Bondar {\it et al.\/},
Phys. Lett. {\bf B356}, 8 (1995).

\bibitem{kr66}
D. S. Koltun and A. Reitan, 
Phys. Rev. {\bf 141}, 1413 (1966).

\bibitem{ms91}
G. A. Miller and P. U. Sauer, 
Phys. Rev. C {\bf 44}, R1725 (1991).

\bibitem{nis92} 
J.A. Niskanen,  Phys. Lett. {\bf B289}, 227 (1992) 

\bibitem{wei66}
S. Weinberg, Phys. Rev. Lett. {\bf 17}, 616 (1966).

\bibitem{hoe83}
G. H\"{o}hler, 
in {\it Pion - Nucleon Scattering}, ed. K.H. Hellwege,
Landolt-B\"{o}rnstein, New Series, Group I, vol. 9 b2., 
(Springer-Verlag, New York, 1983).

\bibitem{lr93}
T.-S. H. Lee and D. O. Riska, 
Phys. Rev. Lett. {\bf 70}, 2237 (1993).

\bibitem{br92}
P. G. Blunden and D. O. Riska, 
Nucl. Phys. {\bf A536}, 697 (1992); 
K. Tsushima, D. O. Riska and P. G. Blunden, 
Nucl. Phys. {\bf A559}, 543 (1993).

\bibitem{hmg94}
C. J. Horowitz, H. O. Meyer and D. K. Griegel, 
Phys. Rev. C {\bf 49}, 1337 (1994).

\bibitem{ho95}
E. Hern\'{a}ndez and E. Oset, 
Phys. Lett. {\bf B350}, 158 (1995).

\bibitem{ham67}
G. Hamilton, High Energy Physics, {\it ed. \/} 
E. H. S. Burhop, Vol. 1, p. 194 
(Academic Press, New York, 1967). 

\bibitem{hanetal95}
C. Hanhart, J. Haidenbauer, A. Reuber, 
C. Sch\"{u}tz and J. Speth,  Phys. Lett. 
{\bf B358}, 21 (1995).

\bibitem{gl84}
J. Gasser and H. Leutwyler, 
Ann. Phys. {\bf 158}, 142 (1984).

\bibitem{bkm95}
For a review, see e.g.,
V. Bernard, N. Kaiser and Ulf-G. Meissner, 
Int. J. Mod. Phys. {\bf E4}, 193 (1995)

\bibitem{jm91}
E. Jenkins and A. V. Manohar, 
Phys. Lett. {\bf B255}, 558 (1991).

\bibitem{wei90} 
S. Weinberg, 
Phys. Lett. {\bf B251}, 288 (1990);
Nucl. Phys. {\bf B363}, 3 (1991);
Phys. Lett. {\bf B295}, 114 (1992).

\bibitem{kol92}
U. van Kolck, thesis, University of Texas at Austin,
(1992);
C. Ordonez, L. Ray and U. van Kolck,
Phys. Rev. Letters, {\bf 72}, 1982 (1994).

\bibitem{pmr93}
T. S. Park, D.-P. Min and M. Rho, 
Phys. Repts. {\bf 233}, 341 (1993).

\bibitem{ptk94}
T. S. Park, I. S. Towner and K. Kubodera, 
Nucl. Phys. {\bf A579}, 381 (1994).

\bibitem{pmr95}
T. S. Park, D.-P. Min and M. Rho, 
Phys. Rev. Lett. {\bf 74}, 4153 (1995); 
``Chiral Lagrangian approach to exchange 
vector currents in nuclei",
preprint SNUTP 95-043 (nucl-th/9505017), 1995.

\bibitem{cfmv95}
T.D. Cohen, J.L. Friar, G.A. Miller and U. van Kolck,
preprint: 
``The $pp \rightarrow pp\pi^0$ reaction near threshold:
A chiral power counting approach,"
nucl-th/9512036

\bibitem{geo84}
see e.g., H. Georgi, {\it Weak Interactions and 
Modern Particle Theory}, (Benjamin, 1984)

\bibitem{com1}
This is the form used by Bernard et al.
\cite{bkm95}.
Another commonly used parameterization
is the ``exponential form",
see e.g. \cite{pmr93,geo84} 
where $U(x)=\exp[i\bbox{\tau} \!\cdot\! 
\bbox{\pi}(x)/f_\pi]$.

\bibitem{com2}
In practical calculations 
we will choose the nucleon rest frame 
$v_\mu = (1, {\bf 0})$
in which case Eq.(\ref{N}) corresponds to 
the standard non-relativistic reduction of 
a spinor into upper and lower
components and the covariant spin operator
of Eq.(\ref{S})  
is simply given by 
$S = (0, \frac12 \bbox{\sigma})$.

\bibitem{com3}
In Eq.(11a) the sum over $A$ runs over 
the possible combinations of 
$\gamma$- and $\bbox{\tau}$- matrices: 
$\Gamma_{S}^{S} = 1$, 
$\Gamma_{S}^{V} = \bbox{\tau}$, 
$\Gamma_{V}^{S} = S_\mu$, 
and $\Gamma_{V}^{V} = S_\mu \bbox{\tau}$. 
However, because of the Fermi statistics 
(Fierz rearrangement), 
only two of the four coupling constants $C_A$ 
are independent.

\bibitem{bkm93}
V. Bernard, N. Kaiser and Ulf-G. Meissner, 
Phys. Lett. {\bf B309}, 421 (1993).

\bibitem{gls91}
J. Gasser, H. Leutwyler and M. E. Sainio, 
Phys. Lett. {\bf B253}, 252, 260 (1991).

\bibitem{koc86}
R. Koch, Nucl. Phys. {\bf A448}, 707 (1986). 

\bibitem{com4}
We may remark en passant
that this ``standard value"
for $\lambda_1$ determined from $a^+$,
is numerically close 
to the contribution of the crossed Born term,
which in HFF is
grouped with the $c_2$ term in ${\cal L}^{(1)}$
as $\frac{g_A^2}{8 m_N}$.
This remnant of the s-wave pion nucleon 
crossed Born term in HFF,
appears in the second term in Eq.(\ref{eq:kappa}),
and 
gives a value of $a^+$, 
$$
%\frac{m_\pi^2}{f_\pi^2} 
%\cdot \frac{g_A^2}{8 m_N} 
%\approx 0.47 {\rm GeV}^{-1}.
a^+ = - \frac{1}{2 \pi} \cdot
\left( 1+\frac{m_{\pi}}{m_N} \right) \cdot
\frac{m_\pi^2}{f_\pi^2} 
\cdot \frac{g_A^2}{8 m_N} 
\approx - 0.009 m_{\pi}^{-1}
$$
%the first term in
%Eq.(\ref{eq:kappa1}), 
compatible with
${a^+}_{exp} = ( - 0.83 \pm 0.38 ) \cdot 10^{-2} {m_\pi}^{-1} $ \cite{hoe83}.

\bibitem{wat52}
K. M. Watson, Phys. Rev. {\bf 88}, 1163 (1952).

\bibitem{hj62}
T. Hamada and I. D. Johnston, 
Nucl. Phys. {\bf 34}, 382 (1962). 

\bibitem{rei68}
R. V. Reid,  Ann. Phys. {\bf 50}, 411 (1968).

\bibitem{cfmv95a}
This type of cancellation has also been noted 
in Ref.\cite{cfmv95}.


\end{references}
\end{document}